\documentstyle[prl,aps,preprint]{revtex}
\tightenlines
\newcommand{\be}[1]{\begin{equation}\label{#1}}
\newcommand{\ee}{\end{equation}}     
\newcommand{\bea}{\begin{eqnarray}}
\newcommand{\eea}{\end{eqnarray}} 
\newcommand{\eq}[1]{Eq.~(\ref{#1})}

\begin{document}  
\draft 
\title{Threshold fragmentation under dipole forces}
\author{Thomas Pattard and Jan M.\ Rost} 
\address{
 Max-Planck-Institute for the Physics of Complex Systems, 
N\"othnitzer Str.\ 38, D-01187 Dresden, Germany\\[3mm]
}
\date{\today}
\maketitle
\begin{abstract} 
The threshold law for $N-$body fragmentation under dipole forces is 
formulated. It  emerges from the
energy dependence of the normalization of the  correlated continuum wave
function for $N$ fragments.  It is shown that the dipole threshold law 
plays a key role in understanding all threshold fragmentation
phenomena since  it links the classical threshold law for long-range Coulomb
interactions to the  statistical law for short-range interactions. Furthermore,
a tunnelling mechanism is identified as the common feature
which occurs for all three classes of interactions, short-range, dipole and
Coulomb.
\draft
\end{abstract}
In 1949 Wigner derived a threshold law for the break-up of two quantum 
particles under short-range and long-range (Coulomb) forces 
\cite{Wig49}.  Using exclusively classical mechanics Wannier 
formulated in 1953 a threshold law for the break-up of a two-electron atom
into three charged particles \cite{Wan53}. Corresponding threshold laws for four
charged fragments have been published in the meantime \cite{Poeal94}.  
More than 15 years after Wigner's paper O'Malley provided the two-body 
threshold law for dipole interactions \cite{OMa65} through the 
analysis of the normalization constant of the continuum final state 
wavefunction in a similar way as Wigner obtained his threshold laws.  
Wigner's as well as Wannier's threshold law have been confirmed by a 
number of experiments \cite{experiments}.  
This is also true for the statistical 
threshold law for break-up into multiple fragments under short-range 
forces: Derived from simply counting the available states of free 
motion in the continuum at the respective energy it has been used 
to  successfully 
interprete experimental results and it has been shown to be 
compatible with Wigner's law for short-range forces \cite{Rau84}.

Summarizing the situation one can say that three types of threshold 
laws, Wigner's, Wannier's and the statistical one, have been derived 
by different means for different situations.  Here, we will show that 
a connection between these threshold laws exists: It is the
threshold law for N-body break-up under dipole forces.  We 
will derive it in the following by generalizing O'Malley's two-body 
approach to an arbitrary number of particles.  This allows us to 
formulate the threshold law for $N-$body break-up under dipole 
interactions.  Moreover, for short-range interactions the statistical 
law is directly recovered from the general dipole threshold law.  
Finally, we discuss a semiclassical tunnelling interpretation 
which provides insight into the mechanism of 
threshold fragmentation on the one hand side and clarifies the 
connection with the (semi-)\-classical threshold law for long-range 
Coulomb forces.

Our starting point is the N-particle Schr\"odinger equation in 
hyperspherical coordinates $(r,\vec\omega)$, where $r$ represents the 
hyperradius $r = (\sum_{i}\vec  r_{i}^{2})^{1/2}$.  The $\vec r_{i}$ are 
mass scaled Jacobi coordinates and the set of angles $\vec\omega$ 
denotes the usual geometric directions of the $\vec r_{i}$ in 
configuration space as well as the so called hyperangles which describe the 
relative lengths $r_{i}= |\vec r_i|$, e.g., $\tan\omega_{1} = r_{1}/r_{2}
\cite{Lou60}$.  
Writing the total wavefunction as
\be{wavef}
\psi(r,\vec\omega) = r^{(D-1)/2}\Psi(r,\vec\omega)
\ee
with $D = 3 N - 3$,
we obtain the Schr\"odinger equation
\be{schrod}
\left(\frac{\partial^{2}}{\partial r^{2}}
-\frac{\Lambda^2-J_D}{r^2} - V_{LR}  - 
V_{SR}+k^2\right)\Psi(r,\vec\omega)= 0\,,
\ee
where the energy has been expressed through the wavenumber 
$k = (2mE)^{1/2}/\hbar$. The Jacobian factor $J_D =
\textstyle\frac 14(D-1)(D-3)$ is  a consequence of the transformation
\eq{wavef} and $\Lambda^2$ is the Laplace operator on the
$D$-dimensional unit sphere \cite{Lou60}. In \eq{schrod} 
we have split the potential into the long-range  part $V_{LR}(r,\vec\omega) = 
2C(\vec\omega)/r^{2}$  and the short-range part defined by 
the property  
\be{lim}
\lim_{k\to 0} k^{-2}V_{SR}(r/k,\vec\omega) = 0
\ee
for all finite $r$.
Since we are interested in the threshold region $k\to 0$ we can 
scale  $r = R/k$  and divide \eq{schrod}  by $k^2$.  
In the limit $k\to 0$
the short-range potential vanishes due to \eq{lim}.  Hence, the 
angular problem in $\vec\omega$ becomes independent of $r$ and the wavefunction
can be written as $\Psi(R,\vec\omega)= \sum_j u_j(R)\Phi_j(\vec\omega)$, where
$\Phi_j$ is an eigenfunction of 
\be{Lambda}
(\Lambda^{2} + 2C(\vec\omega) - \lambda_{j})\Phi_{j}(\vec\omega) = 0
\ee
with eigenvalue $\lambda_j$.
The remaining radial problem represents  the differential equation for 
a Bessel function if we  insert the eigenvalue $\lambda_{j}$  
(for reasons of clarity we continue using unscaled coordinates):
\be{radial}
\left(\frac{\partial^{2}}{\partial r^{2}} 
-\frac{\nu_j^2-\textstyle\frac{1}{4}}{r^{2}}+k^2\right)u_{j}(r).
 \ee
From the effective radial potential
 \be{dipol}
V_{eff}(r) = \frac{ \frac 14
(D-1)(D-3)+\lambda_j}{r^{2}}\equiv
\frac{\nu_j^2-\textstyle\frac{1}{4}}{r^{2}}
\ee
follows
\be{stren}
\nu_j  = [\textstyle\frac 14(D-2)^{2}+\lambda_{j}]^{1/2}
\ee
for  $\nu_j>0$.   
If $(D-2)^{2}<-4\lambda_{j}$ then 
$\nu_j = i\bar\nu_j$ becomes imaginary.
In this case the dipole potential is 
so attractive due to the negative value of $\lambda_j$ that it overcomes 
the repulsive $D-$dependent part.

With \eq{radial} we have reduced the problem of finding the 
threshold  behavior of multi-particle break-up under dipole forces 
to the corresponding problem for two particles solved by O'Malley 
\cite{OMa65}. The only difference is that the strength of the dipole 
potential \eq{stren}  
depends now on  the 
dimension of the problem, D = (3N-3), i.e.\ the number of particles 
$N$, and on the dynamics in the other than radial degrees of freedom 
through the eigenvalue $\lambda_{j}$. 
The solution $u_j(r)$ to \eq{radial} is a linear combination of
Bessel functions $J_{\pm \nu_j}(kr)$. 

The energy dependence of the threshold cross section can be extracted 
from the energy dependence of the normalization constant of $u_j(r)$.  
According to \cite{OMa65} it is given by
\begin{mathletters}
\label{thres}
\bea
\sigma_{j}\propto &k^{2\nu_j},\mbox{\hspace{49mm}} &\nu_j>0 \label{thresa}\\
\bar\sigma_{j}
\propto &[\sinh^2(\bar\nu_j\pi/2)+\cos^2(\bar\nu_j\ln k +\delta_j)
]^{-1},\,\,\, &\nu_j = i\bar\nu_j\label{thresb}
\eea
\end{mathletters}
From \eq{thres} it is clear that the threshold cross section is 
dominated by the `partial wave' with the lowest eigenvalue 
$\lambda_{0}$ if $\nu_0\ge 0$.  In the case of a net attractive 
dipole-potential $\nu_j = i\bar\nu_j$ the threshold cross section will 
be a superposition $\sigma\propto \sum a_{j}\bar\sigma_{j}$ where the 
$a_{j}$ as well as the $\delta_j$ in \eq{thresb} are determined by the 
short-range part of the potential and all partial waves contribute for 
which the eigenvalue $\lambda_j$ is sufficiently negative to yield an 
imaginary $\nu_j$.  For strong enough attractive dipole forces, the 
threshold cross section will approach a constant value.  However, the 
latter case will be the exception for many-particle systems since the 
repulsive centrifugal barrier $(D-2)^{2}\propto N^{2}$ grows much 
faster with the number $N$ of particles than the eigenvalue 
$\lambda_{0} \propto N$.

The dipole threshold law \eq{thresa} contains also the behavior for short-range 
forces, $C(\vec\omega) = 0$.  Then $\nu_0 = \frac 12(D-2)$ and therefore
with $D = 3N-3$
\be{shortrange}
\sigma_s\propto k^{3N-5}.
\ee
This is exactly the statistical threshold law, derived under the assumption
that the fragmented particles are completely free and that they occupy 
a phase space volume ${\cal S}_E$ only restricted by total energy conservation,
\be{stat}
{\cal S}_E = \int K^{3N-4}\delta(E-K^2/2)dK\int_{\vec\omega} d\vec\omega\propto 
k^{3N-5}.
\ee
Here, we have used again hyperspherical coordinates, this time in
momentum space where the hypermomentum is given by
$K = (\sum_i\vec k_i^2)^{1/2}$
and the set of angles $\vec\omega$ refers
to the Jacobi momenta $\vec k_i$ in the
same way as for the coordinates, described above. The statistical threshold
law states that the cross section close to threshold varies according to the
available final states in the continuum given by their phase space volume
$\sigma_s\propto {\cal S}_E$ with the same energy dependence 
as \eq{shortrange}. Of course, Wigner's law $\sigma \propto \sqrt E$ for 
$N = 2$ is reproduced by \eq{shortrange}. 

It seems that the other extreme of interaction, namely charged
fragments which exert mutual forces through the Coulomb potential 
$V_{LR} = 2C(r,\vec\omega)/r$ even at very
large distances, is also compatible with the dipole law of \eq{thresa} since
Wannier's threshold law for this case is again a power law $\sigma \propto
k^\beta$. The exponent for
two escaping electrons and the remaining  (charged) core of the atom, e.g.,
may be expressed in the form \cite{Ros98}
\be{wanexp}
\beta = \frac{1}{4}
\left( \left(1 + \frac 8{C_*}\frac{d^2C}{d\omega^2}\right)^{1/2}-1\right)\,,
\ee
where $\tan\omega = r_1/r_2$, the ratio between
the two electron-nucleus distances,
$\omega_* = \pi/4$ and $C_*= C(\omega_*)$.
However, this similarity is misleading for two reasons. 
Firstly, Wannier's law is purely
classical. It contains the stability properties of a single 
classical orbit (with $\vec r_1 = - \vec r_2$, denoted as '*' in \eq{wanexp}). 
Secondly, 
the radial Coulomb potential for this orbit is {\em attractive}. 
However, the power law \eq{thresa} belongs to an effectively {\em repulsive} 
dipole  potential. 

On the other hand, the threshold  behavior of  fragmentation 
under repulsive Coulomb forces is an exponential law rather than a 
power law, e.g.\ for electron detachment of 
negative ions by electron impact \cite{expdet}. There, as a result 
of the (semi)-classical tunnelling process under the 
repulsive barrier, the dominant energy variation
near threshold is given by a  Gamow factor 
\be{gamow}
\sigma\propto \exp[-2\Gamma(k)],
\ee
where $\Gamma$ is the imaginary tunnelling action,
\be{action}
\Gamma \equiv i\phi
=  \int(-p^2)^{1/2}dr = \int_{r_i}^{r_t} (-k^2+2C_*/r)^{1/2} dr.
\ee
This tunnelling process leads through the energy scaling of the
homogeneous Coulomb potential to
$\sigma \propto \exp(-a/k)$. 

Nevertheless, the qualitatively  different
threshold laws for  repulsive Coulomb and dipole/short-range interactions 
may be semiclassically interpreted with the same tunnelling mechanism
\eq{gamow}. In the dipole case  the tunnelling action is given by
\be{dipolact}
\Gamma = \int_{r_i}^{r_t}[-k^2+(\nu^2_j-\textstyle\frac 14)/r^2]^{1/2} dr
\stackrel{{k\to 0}}{\longrightarrow}
-\nu_j\ln k,
\ee
which is logarithmically divergent for $k\to 0$. Hence, despite
the exponential form of the Gamow factor \eq{gamow}, inserting 
the action from \eq{dipolact} leads exactly to the power law of \eq{thresa}.
We conclude that  for most interactions, the threshold mechanism can be
interpreted as a tunnelling process.
For two fragments and higher angular momentum, the tunnelling mechanism
has already been proposed by Rau \cite{Rau84}. However, as shown here,
it actually holds
for {\em all} short-range forces,
and for {\em repulsive}
dipole and Coulomb forces. The only exception are attractive dipole forces
with a relatively complicated threshold behavior \eq{thresb} and attractive
Coulomb forces with the well known, classically derived, power law behavior.
A systematic classification of threshold laws according to the nature of the
respective  
interaction is presented in Table \ref{tableI}.  However, one should 
keep in  mind that within the enormous size of the parameter space 
covered by Table I, there are  always exceptional cases \cite{Ihra}.

In summary we have provided the threshold law for fragmentation of 
$N$ particles under dipole forces. Furthermore, it  has been shown 
that this  threshold law links the statistical law 
for short-range forces to the corresponding law for long-range repulsive
Coulomb forces by a semiclassical tunnelling mechanism.  Overlooking the 
threshold fragmentation for all types of interactions  (Table I) a relatively
simple principle emerges which governs this fragmentation: the balance
between potential and kinetic energy for large interparticle 
distances, i.e.\ for large hyperradius $r$.  The quantum mechanical 
kinetic energy scales as $r^{-2}$ and has the corresponding, dipole-like,
long-range characteristics.  Hence,  for short-range potentials the 
kinetic energy dominates the threshold behavior. Indeed,  a 
statistical approach, counting simply available  states of free 
particles, is sufficient to describe this situation, see \eq{stat}.
Although the 
$r^{-2}$ behavior of the kinetic energy is intrinsically quantum 
mechanical, the $\hbar$-dependence  of the  kinetic energy can be 
scaled out in the absence of a potential and the  threshold problem 
can be solved semiclassically by a tunnelling trajectory (\eq{dipolact}).
The other extreme is the Coulomb potential. With its $r^{-1}$  range 
it  reaches further than the kinetic energy. Hence, the 
threshold behavior is decided by  properties of the potential  only 
(essentially its relative curvature, see \eq{wanexp}).  
Finally, the  subtle case of a
dipole potential  $r^{-2}$ is left. 
Here, both parts, kinetic and potential 
energy, contribute on the same footing. 
Consequently, the threshold law for $N-$body 
break-up as it has been derived in \eq{thres}
exhibits a  complicated behavior.  However, since 
the dipole interaction  `interpolates', roughly speaking,
between short-range and Coulomb potentials,  understanding  its
threshold dynamics  allows one to  solve  the general problem
of $N-$ body threshold fragmentation under arbitrary forces,
originally formulated by Wigner for two particles  \cite{Wig49}.

It is a pleasure to dedicate this article to Martin Gutzwiller. His work on
semiclassical theory has been a great inspiration, even more has he himself
been inspiring for all of us who have been lucky enough to exchange ideas
with him.
\begin{table} 
\caption{\label{tableI} Overview of the threshold laws for N-body
fragmentation under different interactions}
\begin{tabular}{l|c|l|c|l}
interaction & type    & energy dependence  of& equation  &mechanism \\
                    &        & threshold     cross section & (see text)     
\\\tableline
short range, &&&\\
$V (r\gg 1) \propto r^{-\alpha}, \alpha > 2$& &power law
& (\ref{shortrange})& semiclassical (tunnelling) \\\tableline
dipole, & repulsive & power law &
(\ref{thresa})& semiclassical\tablenote{includes the quantum eigenvalue of the
angular equation \eq{Lambda}, see text.} (tunnelling)
\\
 $V (r\gg 1)
\propto r^{-2}$ & attractive & oscillating  &(\ref{thresb}) & 
quantum\\\tableline
 Coulomb,  & repulsive & exponential law &
(\ref{gamow}) & semiclassical (tunnelling) \\ 
$V (r\gg 1) \propto r^{-1}$ & attractive & power law & (\ref{wanexp}) &
classical
\end{tabular}
\end{table}

\end{document}